\documentclass{article}
\usepackage{geometry}
\usepackage[fontsize=12pt]{fontsize}
\usepackage{setspace}
\usepackage{mathrsfs,amsmath}
\usepackage{amssymb,amsfonts}
\usepackage{hyperref}
\usepackage{subcaption}
\usepackage{graphicx}
\usepackage{comment}
\DeclareMathOperator{\sinc}{sinc}
\DeclareMathOperator{\asinc}{asinc}
\DeclareMathOperator{\rect}{rect}
\DeclareMathOperator{\Sa}{Sa}
\DeclareMathOperator{\D}{D}
\newcommand{\doi}{doi:}
\renewcommand*{\doi}[1]{\href{http://dx.doi.org/#1}{doi: #1}}
\begin{document}
\onehalfspacing 
\begin{center}
\textbf{von MISES TAPERING: A NEW CIRCULAR WINDOWING}
\end{center}
\noindent
\textbf{H. M. de Oliveira}\\
Federal University of Pernambuco, Statistics Department, Brazil.\\ \url{https://orcid.org/0000-0002-6843-0635}

\noindent
\textbf{R. J. Cintra}\\
Federal University of Pernambuco, Statistics Department, Brazil.\\ 
\url{https://orcid.org/0000-0002-4579-6757}

\noindent
\vspace{0.25 cm}

\noindent
\textbf{ABSTRACT:}
Discrete and continuous standard windowing are revisited and a a new taper is introduced, which is derived from the normal circular distribution by von Mises. Both the continuous-time and the discrete-time windows are considered, and their spectra obtained. A brief comparison with further classical window families is performed in terms of their properties in the spectral domain. These windows can be used in spectral analysis, and in particular, in the design of FIR (finite impulse response) filters as  an alternative to the Kaiser window\footnote{part of this paper was presented at the SBrT, Brazil, doi 10.14209/SBRT.2018.179}.\\
\textbf{KEYWORDS:}
von Mises, tapering function, circular distributions, FIR design.
\section{Introduction}
Due to the fact that  many signals present a quasi-periodic nature, the signal processing techniques developed for real variables in the real line may not be appropriate. For circular data \cite{Berens}, \cite{Damien}, it makes no sense to use the sample mean, usually adopted to the data line as a measure of centrality. Circular measurements occur in many areas \cite{Jammalamadaka}, such as chronobiology \cite{Karp-Boss}, economy \cite{Dalkir}, geography \cite{Clark_Burt}, medical (circadian therapy \cite{Kirst}, epidemiology \cite{Gao}...), geology \cite{Watson}, \cite{Prevot}, meteorology \cite{Coles}, acoustic scatter \cite{Jenison} and particularly in signals with some cyclic structure (GPS navigation \cite{Luo}, characterization of oriented textures \cite{Costa}, discrete-time signal processing and over finite fields). Even in political analysis \cite{Gill}. Probability distributions have been successfully used for several purposes: for example, the beta distribution was used in wavelet construction \cite{de_Oliveira}. Here, the von Mises distribution is used in the design of tapers. Tools such as rose diagram \cite{Izbicki} allow rich graphical interpretation. Circular properties for random signals is the main focus here. The uniform distribution of an angle $\phi$, circular in the range $[0,2\pi]$, is given by:
\begin{equation}
f_1(\phi):=\frac{1}{2 \pi}\mathbb{I}_{[0,2\pi]}(\phi),
\end{equation}
where $\mathbb{I}_A(.)$ is the indicator function of the interval $A \subset \mathbb{R}$. It is denoted by $\phi \sim \mathcal{U}(0,2\pi)$. Another very relevant circular distribution is the normal circular distribution, introduced in 1918 by von Mises, defined in the interval $[0,2\pi]$ and denoted by $\phi \sim \mathcal{VM}(\phi_0,\beta)$.
\begin{equation}
f_2(\phi):=\frac{1}{2 \pi I_0(\beta)}e^{\beta \cos(\phi - \phi_0)},
\end{equation}
where $\beta \geq 0$ and $I_0(.)$ is the zero-order modified Bessel function of the First Kind \cite{Abramowitz} (not to be confused with the indicator function), i.e. 
\begin{equation}
I_0(z):=\frac{1}{\pi} \int_{0}^{\pi}e^{z.\cos\theta}d\theta=\sum_{n=0}^{+\infty}\frac{(z/2)^{2n}}{{n!}^2}.
\end{equation}
\begin{figure}[ht]
\centering
\includegraphics[scale=0.2]{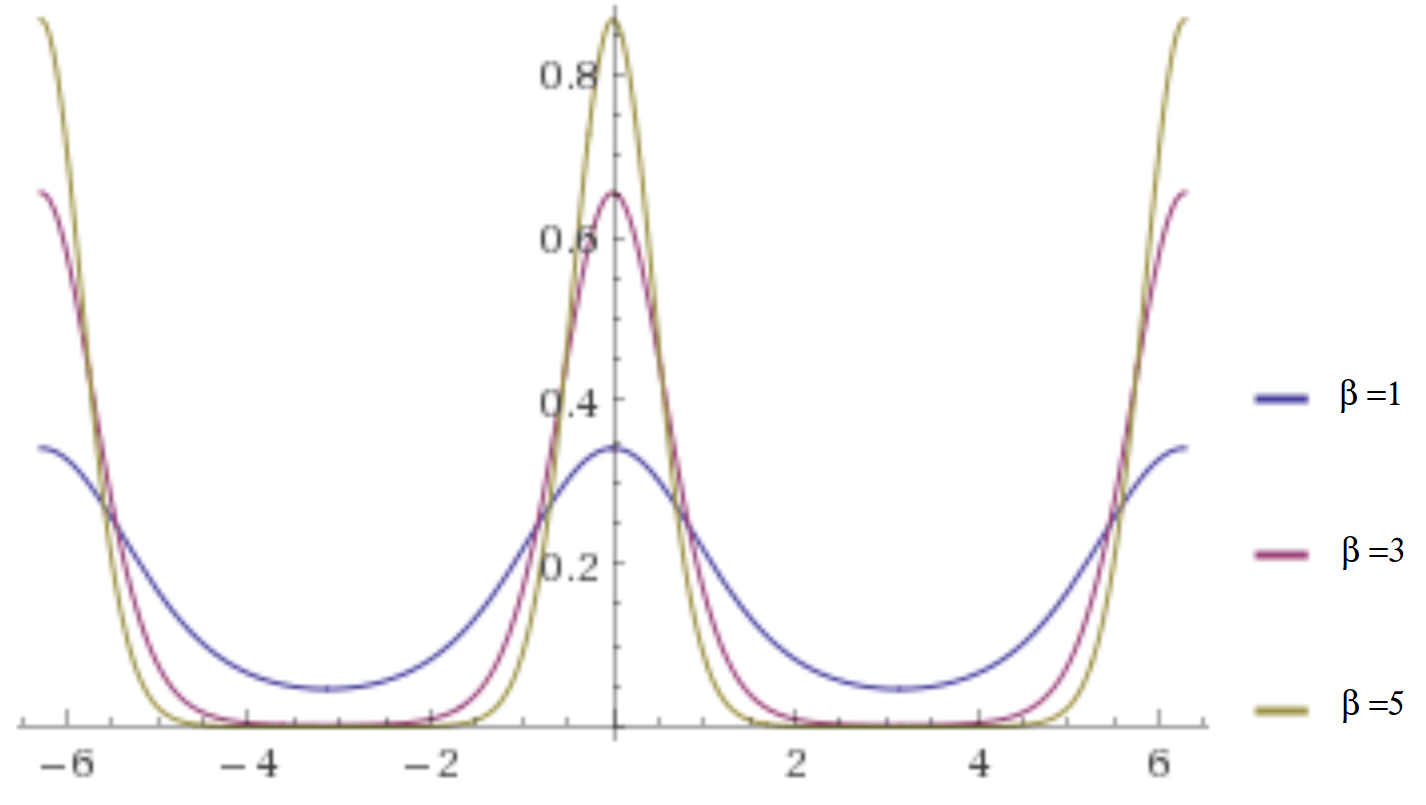}
\caption{\small{Periodic extension of the von Mises distribution with zero-mean for several parameter values:  $\beta=$ 1,3,5. Note that the support of the density is confined to $[-\pi,\pi]$.}}
\label{fig:vonMises}
\end{figure}
This probability density dominates in current analysis of circular data because it is flexible with regard to the effect of parameters. In a standard notation,
\begin{equation}
f(x| \mu, \kappa):= \frac{e^{\kappa \cos(x-\mu)}}{2\pi I_0(\kappa)}.\mathbb{I}_{[-\pi,\pi]}.
\end{equation}
Standardized distribution support is $[-\pi,\pi]$ and the mean, mode and median values are $\mu$. The parameter $\kappa$ plays a role connected to variance, being $\sigma ^2 \approx 1/\kappa$.
\begin{equation*}
\mathbb{E}(X)=\mu ~\textnormal{and}~\mathbb{V}ar(X)=1-\frac{I_1(\kappa)}{I_0(\kappa)}.
\end{equation*}
Two limiting behaviors can be observed:
\begin{itemize}
\item
\begin{equation}
\lim_{\kappa \rightarrow 0} f(x|\mu,\kappa)=\frac{1}{2\pi}\rect \left ( \frac{x}{2\pi} \right ),
\end{equation}
where $\rect(x):= 
\left\{\begin{matrix}
1 & \textnormal{if }|x| \leq 1/2\\ 
0 & \textnormal{otherwise.}
\end{matrix}\right.$ is the gate function, and therefore 
\begin{equation}
\lim_{\kappa \rightarrow 0} \mathcal{VM}(\mu,\kappa) \sim \mathcal{U}(-\pi,\pi).
\end{equation}
\item
\begin{equation}
\lim_{\kappa \rightarrow +\infty } f(x|\mu,\kappa)=\frac{1}{\sqrt{2 \pi \sigma ^2}} e^{-\frac{(x-\mu)^2}{2 \sigma ^2}},
\end{equation}
where $\sigma ^2:=1/\kappa$, and therefore ($\mathcal{N}$ stands for the normal distribution)
\begin{equation}
\lim_{\kappa \rightarrow +\infty} \mathcal{VM}(\mu,\kappa) \sim \mathcal{N}(\mu,\frac{1}{\kappa}).
\end{equation}
\end{itemize}
Hence the reason why this distribution is known as the \textit{circular normal distribution}. The von Mises distribution ($\mathcal{VM}$) is considered to be a circular distribution having two parameters and it corresponds to as a natural analogue of the the Normal distribution on the real line. Maximum entropy distributions are outstanding probability distributions, because maximizing entropy minimizes the amount of prior information built into the distribution. Furthermore, many physical systems tend to move towards maximal entropy configurations over time. This encompasses distributions such as uniform, normal, exponential, beta...
\begin{figure}[ht]
\centering
\includegraphics[scale=0.2]{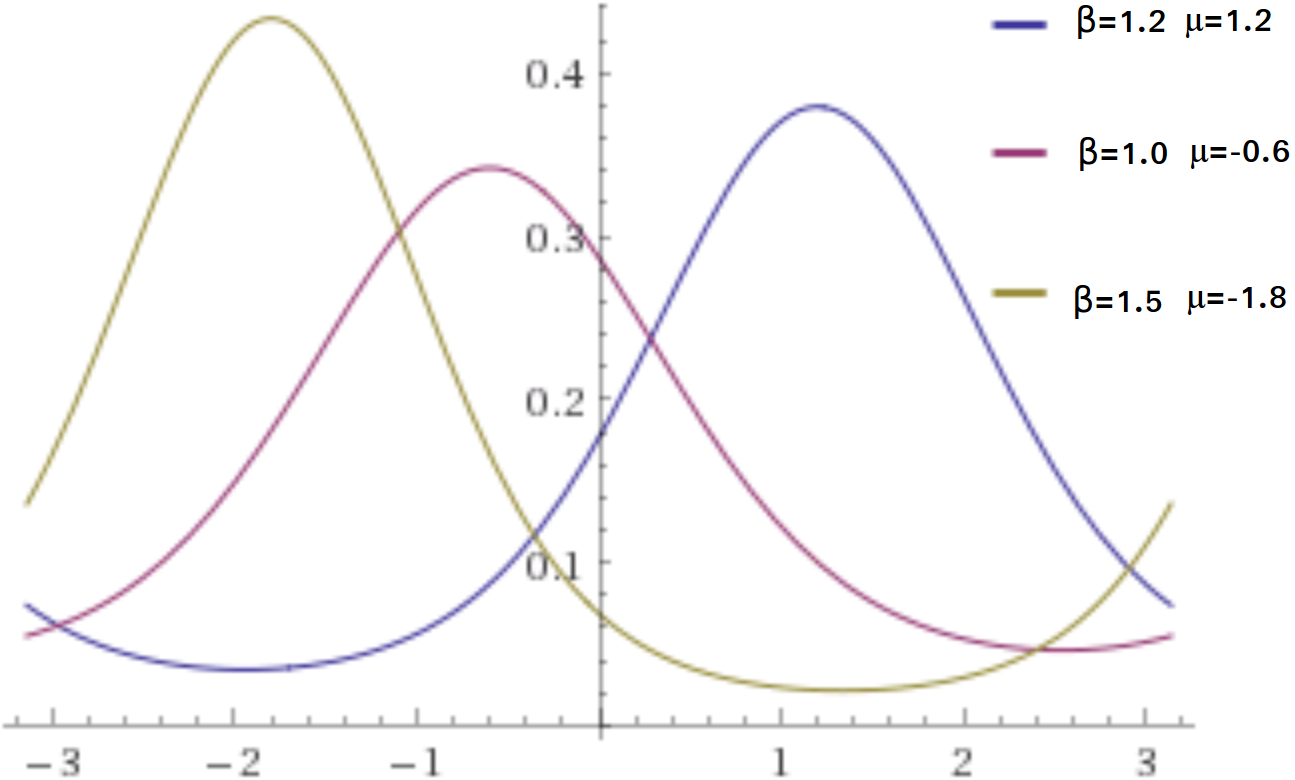}
\caption{\small{Circular behavior of the von Mises distribution plotted for different mean values (1.2, -0.6 and -1.8). The cyclical feature of the distribution is shown outside $[-\pi,\pi]$.}}
\label{fig:ciclicvonMises}
\end{figure}
\\
The von Mises distribution achieves the maximum entropy for circular data when the first circular moment is specified \cite{Jammalamadaka}. The corresponding cumulative distribution function (CDF) is expressed by
\begin{equation}
F_X(x| \mu,\kappa)=\frac{1}{2 \pi}\sum_{n=-\infty }^{+\infty }\frac{I_{|n|}(\kappa)}{I_0(\kappa)}\left ( x-|n| \right ).\Sa\left ( n(x-\mu) \right ),
\end{equation}
where $\Sa(x):=\sin(x)/x, ~x\neq 0$ is the well-known sample function \cite{Lathi}. Through a simple random variable transformation, the distribution support can be modified to an interval defined between two integers:
\begin{equation}
f_{X_1}(x):=\frac{e^{\beta.\cos \left ( \frac{2\pi}{N}x \right ) }}{NI_0(\beta)},~\textnormal{circular in } 0\leq x\leq N.
\end{equation}
Another closely related continuous distribution (with a minimal - but relevant difference) is:
\begin{equation}
f_{X_2}(x):=\frac{e^{\beta.\cos \left ( \frac{\pi}{N}x \right ) }}{NI_0(\beta)},~\textnormal{circular in } 0\leq x\leq N.
\end{equation}
This distribution has a circular pattern as best illustrated in Figure ~\ref{fig:ciclicvonMises}. Decaying pulses for constraining the signal support play a key role in a large number of domains, including: tapers \cite{Durrani}, linear networks (filtering \cite{Hayes}, inter-symbolic interference control \cite{Lathi}), wavelets \cite{deOliveira}, time series, Fourier transform spectroscopy \cite{Naylor} ... \\
Digital filters can be characterized by their impulse response $h[n]$ or their transfer function $H(z)$, related by the $z$-transform \cite{Hayes}:
\begin{equation}
H(z)=\mathscr{Z}(h[n]):=\sum_{n=-\infty }^{+\infty }h[n]z^{-n}.
\end{equation}
The frequency response can be evaluated by setting $z=e^{j\omega}$, yielding
\begin{equation}
H(e^{j\omega})=\sum_{n=-\infty }^{+\infty }h[n]e^{-jn\omega}.
\label{eq:infinite_series}
\end{equation}
The window method consists of simply ``windowing'' a theoretically ideal filter impulse response $h[n]$ by some suitably chosen apodization function $w[n]$, yielding
\begin{equation}
h_w[n]:=w[n].h[n], ~~~~~~n \in \mathbb{Z}.
\end{equation}
This results in a truncation of the infinite series referred to in Eqn.\eqref{eq:infinite_series} (a FIR), i.e.,
\begin{equation}
H_w(e^{j\omega})=\sum_{n=-N/2 }^{N/2} w[n].h[n]e^{-jn\omega}.
\end{equation}
It can be found in the literature numerous articles dealing with the application of windows in FIR filter designs \cite{Smith} among others.
For example, for the ideal lowpass filter (LPF), the impulse response is $\frac{1}{2}\sinc(\frac{n}{2})~~n \in \mathbb{Z}$.
\section{Tapering: Standard Windows}
\begin{figure}[ht]
\centering
{\includegraphics[width=0.48\columnwidth, height=2.3cm]{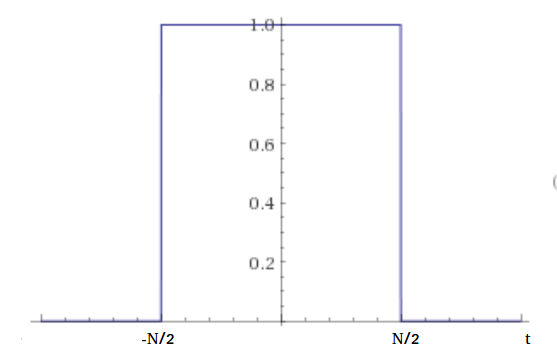}}
{\includegraphics[width=0.48\columnwidth, height=2.3cm]{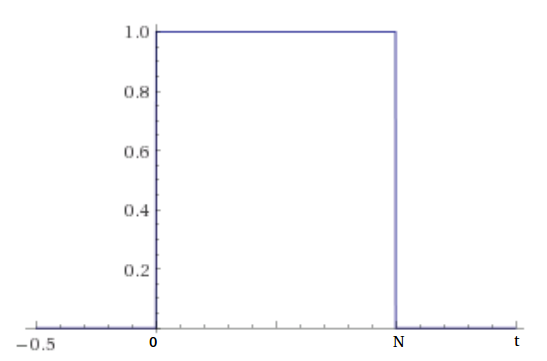}}
{\includegraphics[width=0.48\columnwidth]{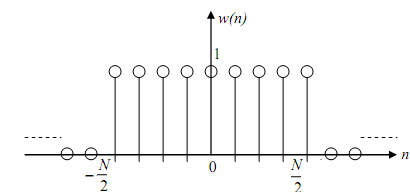}}
{\includegraphics[width=0.48\columnwidth]{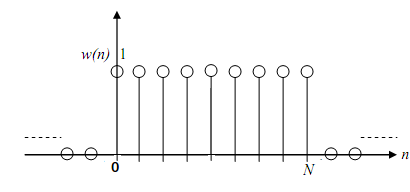}}
\caption{\small{Rectangular windows with length $N$ (4 types): a) continuous, b) continuous causal, c) discrete, d) discrete causal windows. }}
\label{fig:rect_windows}
\end{figure}

Here we review some of the continuous and discrete windows (also known as a apodization function) used in signal processing (spectrum analysis \cite{Nuttall}, \cite{Durrani}), antenna array design \cite{Van_Veen},  characterization of oriented textures \cite{Costa}, image warping and filtering (FIR filter design \cite{Shayesteh}, \cite{Hayes}). Although discrete windows are more common, some studies addressing continuous windows \cite{Theubl}, \cite{Geckinli}, besides their application in short-time Fourier transforms. Among the most frequently used windows, it is worth mentioning: \textit{Rectangular, Bartlett, cosine-tip, Hamming, Hanning, Blackman, Lanczos, Kaiser, modified Kaiser, de la Vallé-Pousin, Poisson, Saram"aki} \cite{Saramaki}, \textit{Dolph-Chebyshev...} (non-exhaustive list \cite{Poularikas}). Tutorials on the subject are available \cite{Gautam}, \cite{Poularikas}, \cite{Harris}. The Figure ~\ref{fig:rect_windows} describes the approaches to the standard window, i.e. the rectangular window, considering four cases: 1) continuous non-causal, 2) continuous causal, 3) discrete non-causal, and 4) discrete causal window. These indexes are used in subscribed windows (time and frequency).
It is worth revisiting the spectra of each of these windows.
\begin{subequations}
\begin{align}
W_{REC;1}(t)&=\rect\left ( \frac{t}{N} \right ),\label{a}\\
W_{REC;2}(t)&=\rect\left ( \frac{t-N/2}{N} \right ).\label{b}
\end{align}
\end{subequations}
In the continuous case, $w_1(t)$ has spectrum given by:
\begin{equation}
W(w):=\mathscr{F}[w(t)]=\int_{-\infty }^{+\infty }w(t)e^{-jwt}dt.
\label{eq:ContinuoustimeFOURIER}
\end{equation}
Indeed $W_{REC;1}(w)=N .\Sa\left ( \frac{wN}{2}  \right )$. Now the spectrum of $w_{REC;2}(t)=\rect\left ( \frac{t-N/2}{N}  \right )$ can be evaluated using the time-shift theorem \cite{Lathi}, $w(t-t_0)\leftrightarrow W(w).e^{-jwt_0}$, resulting in $W_{REC;2}(w)=N .\Sa\left ( \frac{wN}{2}  \right )e^{-jwN/2}$. The corresponding discrete-time windows are:
\begin{subequations}
\begin{align}
w_{REC;3}[n]&=\left\{\begin{matrix}
1 & -N/2 \leq n \leq N/2 \\ 
0 & \textnormal{otherwise.}
\end{matrix}\right.\label{a}\\
w_{REC;4}[n]&=\left\{\begin{matrix}
1 & 0 \leq n \leq N \\ 
0 & \textnormal{otherwise.}
\end{matrix}\right. \label{b}
\end{align}
\end{subequations}
In the case of discrete signals (discrete time), the discrete-time Fourier Transform (DTFT) is used:
\begin{equation}
W\left ( e^{j\omega} \right ):=\sum_{n=-\infty }^{+\infty }w[n].e^{-jn\omega}.
\label{eq:DiscretetimeFOURIER}
\end{equation}
The idea behind the use of windowing is to confine the previous summation. For the window $w_{REC;3}[n]$, we have a spectrum:
\begin{equation}
W_{REC;3}\left ( e^{j\omega} \right )=\sum_{n=-N/2 }^{N/2 }e^{-jn\omega}=\frac{\sin(\frac{N+1}{2}w)}{\sin(\omega/2)}.
\end{equation}
This substantially corresponds to the Dirichlet kernel $\D(\omega):=\frac{\sin\frac{N+1}{2}\omega}{\sin\frac{\omega}{2}}e^{j\frac{\omega}{2}}$ (or periodic sinc function) \cite{Dirichlet}. For the causal discrete rectangular $w_{REC;4}[n]$, 
\begin{equation}
W_{REC;4}\left ( e^{j\omega} \right )=\sum_{n=0}^{N}e^{-jn\omega}=\frac{\sin(\frac{N+1}{2}w)}{\sin(\omega/2)} e^{-j\frac{\omega}{2}N}.
\end{equation}
\\
A less adopted but appealing notation is the \textit{aliased sinc function}
\begin{equation}
\asinc_M(\omega):=\frac{\Sa(M.\omega/2)}{\Sa(\omega/2)}=\frac{\sinc(Mf)}{\sinc(f)}.
\end{equation}
For the discrete causal window, the time shift property for the discrete-time Fourier transform can also be used. Several of the windows of interest can be encompassed taking into account the following definition:
\begin{equation}
w_{\alpha;1}(t):= \left \{ \alpha+(1-\alpha).\cos\left ( \frac{2\pi}{N}t\right ) \right \}.\rect\left ( \frac{t}{N} \right ),
\label{eq:general_alpha}
\end{equation}
The Hanning (raised cosine) window corresponds to $\alpha=0.5$, whereas the standard Hamming window corresponds to  $\alpha=0.54$ \cite{Podder}. In the case of a cosine-tip continuous window ($\alpha=0$), the corresponding window and spectrum are \cite{Geckinli}:
\begin{equation}
w_{\alpha=0;1}:=\cos\left ( \frac{2\pi}{N}t\right ).\rect\left ( \frac{t}{N} \right ),
\end{equation}
and therefore,
\begin{equation}
W_{\alpha=0;1}(w)=\frac{N}{2}.\Sa\left ( \frac{Nw}{2}-\pi  \right )+ \frac{N}{2}.\Sa\left ( \frac{Nw}{2} +\pi \right ).
\end{equation}
In the discrete case,
\begin{equation}
W_{\alpha=0;4}(e^{j\omega})=\frac{1}{2}\left [ \D\left ( \omega-\frac{2\pi}{N} \right )+ \D\left ( \omega+\frac{2\pi}{N} \right )\right ].e^{-j\omega\frac{N}{2}}
\end{equation}
We shall denote alternatively by
\begin{equation}
W_{\alpha=0;3}(e^{j\omega})=\frac{1}{2}\left [ \asinc_{N+1}\left ( \omega-\frac{2\pi}{N} \right )+ \asinc_{N+1}\left ( \omega+\frac{2\pi}{N} \right )\right ].
\label{eq:alpha0}
\end{equation}
For the discrete-time case with arbitrary $\alpha$ (Eqn.\eqref{eq:general_alpha}), we have the following linear combination of spectra:
\begin{equation}
W_{\alpha;3}(e^{j\omega})=\alpha.W_{\alpha =1;3} ( e^{j\omega})+(1-\alpha).W_{\alpha=0;3}(e^{j\omega}).
\end{equation}
The Kaiser window in continuous variable is defined by (non-causal window centered on the origin, and its corresponding causal version)
\begin{subequations}
\begin{align}
w_{KAI;1}(t):=&\frac{I_0\left ( \beta\sqrt{1-\left ( \frac{t}{N/2} \right )^2} \right )}{I_0(\beta))}.\rect\left ( \frac{t}{N} \right ),\label{a}\\
w_{KAI;2}(t):=&
\frac{I_0\left ( \beta\sqrt{1-\left ( \frac{t-N/2}{N/2} \right )^2} \right )}{I_0(\beta))}.\rect\left ( \frac{t-N/2}{N} \right ).\label{b}
\end{align}
\end{subequations}
In the case of discrete versions (those that are used in filter designs), the corresponding versions are \cite{Kaiser}: a) non causal and b) causal, respectively.
\begin{subequations}
\begin{align}
&w_{KAI;3}[n]:=\left\{\begin{matrix}
\frac{I_0\left ( \beta\sqrt{1-\left ( \frac{n}{N/2} \right )^2} \right )}{I_0(\beta))} & 0 \leq n \leq N \\ 
0 & \textnormal{otherwise.}
\end{matrix}\right. \label{a}\\
&w_{KAI;4}[n]:=\left\{\begin{matrix}
\frac{I_0\left ( \beta\sqrt{1-\left ( \frac{n-N/2}{N/2} \right )^2} \right )}{I_0(\beta))} & -N/2 \leq n \leq N/2 \\ 
0 & \textnormal{otherwise.}
\end{matrix}\right. \label{b}
\end{align}
\end{subequations}
The spectrum of discrete Kaiser windows can be evaluated resulting in \cite{Kaiser}:
\begin{equation}
W_{KAI;3}\left ( e^{j\omega} \right )=\frac{N}{I_0(\beta)}\Sa\left ( \sqrt{\left ( \frac{N\omega}{2} \right )^2 - \beta ^2} \right ).
\end{equation}
\section{Introducing the Circular Normal Window}
The proposal here is to use a window (support length $N$)  with shape related to
\begin{equation}
W(t)=K.\frac{e^{\beta.\cos\left ( \frac{\pi}{N}t \right )}} {I_0(\beta)}.\rect\left ( \frac{t}{N} \right ).
\end{equation}
The value of the constant $K$ can be set so that, as in the other classic windows, $w(0)=1$. Thus, for continuous cases (both not causal and causal), one has:
\begin{subequations}
\begin{align}
W_{CIR;1}(t)&=\frac{e^{\beta.\cos\left ( \frac{\pi}{N}t \right )}} {e^\beta}.\rect\left ( \frac{t}{N} \right ),\label{a}\\
W_{CIR;2}(t)&=\frac{e^{\beta.\cos\left ( \frac{\pi}{N}t \right )}} {e^\beta}.\rect\left ( \frac{t-N/2}{N} \right ),\label{b}
\end{align}
\end{subequations}
For the discrete circular windows in time, consider the definitions:
\begin{subequations}
\begin{align}
W_{CIR;3}[n]&=e^{\beta.\left [ \cos \left ( \frac{n\pi}{N} \right )-1 \right ]},~~~~~ |n|\leq N/2.\\
W_{CIR;4}[n]&=e^{\beta.\left [ \cos \left ( \frac{\pi}{N}(n-\frac{N}{2}) \right )-1 \right ]},~~~~~ 0\leq n \leq N.
\end{align}
\end{subequations}
\begin{figure}[ht]
\centering
\begin{subfigure}{0.45\textwidth}
\includegraphics[scale=0.25]{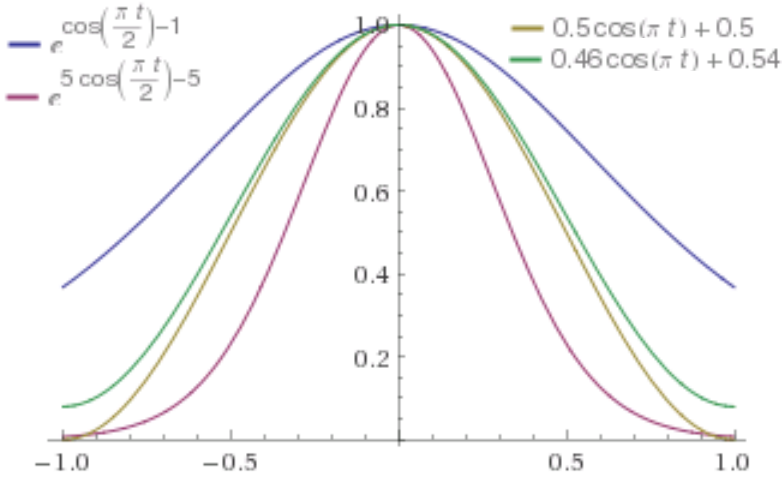}
\captionof{figure}{\small{von Hann, Hamming, circular $\beta=1,~5$.}}
\end{subfigure}%
\hfill
\begin{subfigure}{0.45\textwidth}
\centering
\includegraphics[scale=0.25]{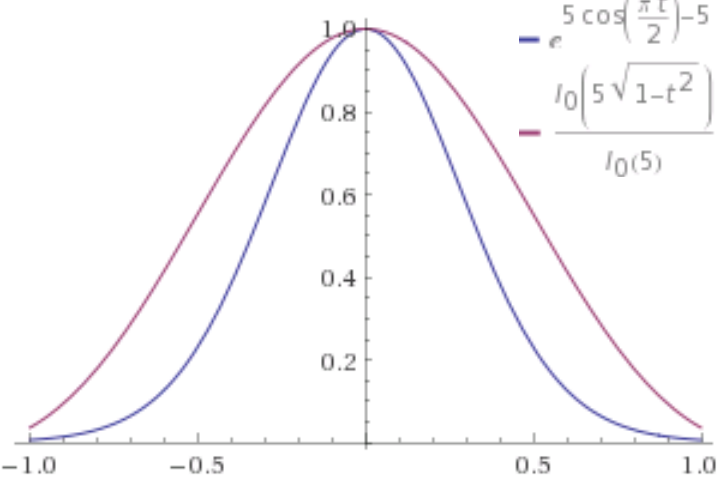}
\captionof{figure}{\small{Kaiser vs normal circular windows: $\beta=5$.}}
\end{subfigure}%
\caption{Shape comparison of different normalized windows for support: $[-1,1]$.}
\label{fig:comparison_Han_Ham_CIR}
\end{figure}
\section{Spectrum of the Normal Circular Window: the Continuous Case}
\begin{figure}[ht]
\centering
\includegraphics[width=9cm,height=4.0cm]{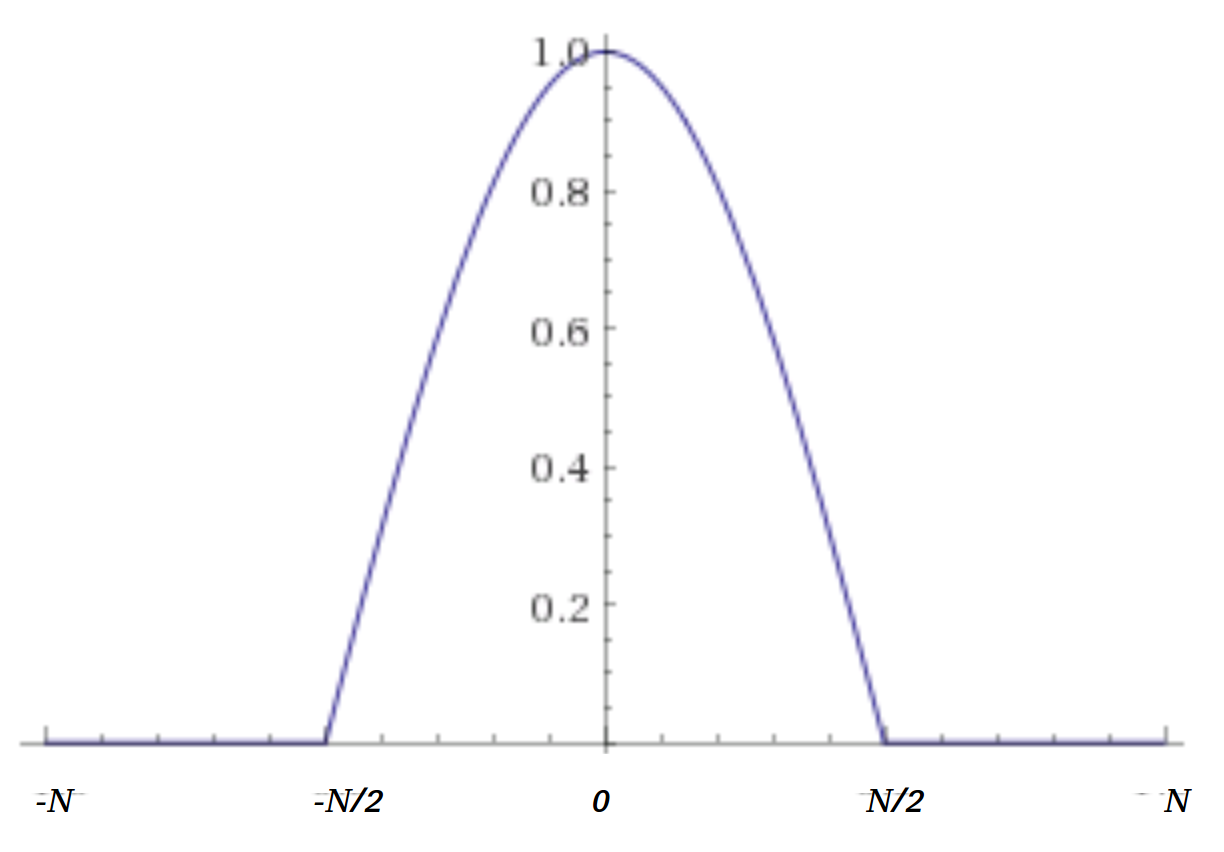}
\caption{\small{Normalized cosine exponent of the exponential function in von Mises window: the (entire) cosine $\cos(\pi t/N)$ is periodic in $[-N,N]$, but the support is confined within $[-N/2,N/2]$ due to the rectangular pulse.}}
\label{fig:cut_by_gate_window}
\end{figure}

In order to evaluate the spectrum of the continuous window introduced in the previous section, we use Eqn.~\eqref{eq:ContinuoustimeFOURIER},
\begin{equation}
W_{CIR;1}(w)=\int_{-N/2}^{N/2}e^{\beta.\left [ \cos \left ( \frac{\pi}{N}t \right )-1 \right ]}e^{-jwt}dt
\end{equation}
The interest function involved in defining the window is $\cos \left ( \frac{\pi}{N}t \right )$, with period $2N$, sketched below in $[-N,N]$.  The rectangular term included in the window is responsible for cutting the window, confining it in the range $[-N/2,N/2]$ as viewed in Figure ~\ref{fig:cut_by_gate_window}. 
MacLaurin's serial development of $e^{\beta.\left [ \cos \left ( \frac{\pi}{N}t \right ) \right ]}$ gives:
\begin{equation}
e^{\beta.\left [ \cos \left ( \frac{\pi}{N}t \right ) \right ]}=\sum_{n=-\infty }^{+\infty }I_{|n|}(\beta)
\cos\left (  \frac{n \pi}{N}t \right ).
\end{equation}
Thus, one obtains:
\begin{equation}
W_{CIR;1}(w)=e^{-\beta}\sum_{n=-\infty }^{+\infty }I_{|n|}(\beta).
\mathscr{F}~\left ( \cos\left (  \frac{n \pi}{N}t \right ).\rect \left ( \frac{t}{N} \right ) \right ).
\end{equation}
From the property of the convolution (\cite{Lathi}), the spectrum sought is:
\begin{equation}
W_{CIR;1}(w)=\frac{1}{2\pi}e^{-\beta}\sum_{n=-\infty }^{+\infty }I_{|n|}(\beta).
\mathscr{F}~\left ( \cos\left (  \frac{n \pi}{N}t \right )\right )* \mathscr{F} ~\left ( \rect \left ( \frac{t}{N} \right ) \right ).
\end{equation}
By evaluating the internal terms of the summation, one come easily to
\begin{equation}
\frac{N}{2\pi}.\pi.\left \{ \delta \left ( w-\frac{n\pi}{N} \right )+\delta \left ( w+\frac{n\pi}{N} \right ) \right \}*\Sa\left ( \frac{wN}{2} \right ),
\end{equation}
where $\delta$(.) is the Dirac impulse \cite{Lathi} and finally,
\begin{equation}
W_{CIR;1}(w)=\frac{N}{2}.e^{-\beta}.\sum_{n=-\infty }^{\infty } I_{|n|}(\beta)
\left \{ \Sa \frac{N}{2} \left ( w-\frac{n\pi}{N} \right ) +\Sa \frac{N}{2} \left ( w+\frac{n\pi}{N} \right )\right \},
\end{equation}
so,
\begin{equation}
W_{CIR;1}(w)=N.e^{-\beta}.\sum_{n=-\infty }^{\infty } I_{|n|}(\beta) \left \{ \Sa \left ( \frac{Nw}{2}-\frac{n\pi}{2} \right ) \right \}.
\label{eq:series}
\end{equation}
This expression is as if a series of reconstitution (with coefficients $c_n$)  of the type:
\begin{equation*}
\sum_{n=-\infty }^{+\infty }c_n.\Sa\left ( \frac{Nw}{2}-\frac{n\pi}{2} \right ).
\end{equation*}
Let us now apply the Shannon-Nyquist-Koteln'kov sampling theorem in the frequency domain, for time-limited signals (\cite{Jerri}, \url{http://ict.open.ac.uk/classics}.
\begin{equation}
F(w)=\frac{w_st_m}{\pi}\sum_{n=-\infty }^{+\infty }F(nw_s)\Sa \left ( wt_m-nt_mw_s \right ).
\end{equation}
The rate $w_s$ must comply with the restriction $w_s  \leq \pi/t_m$, and the choice made is $w_s= \pi/2t_m$, so that the previous equation is:
\begin{equation}
F(w)=\frac{1}{2}\sum_{n=-\infty }^{+\infty }F(\frac{n\pi}{2t_m})\Sa \left (w t_m -\frac{n\pi}{2} \right ).
\end{equation}
Now let us choose the duration $t_m$ to be $t_m:=N/2$ (Figure ~\ref{fig:cut_by_gate_window}).
\begin{equation}
F(w)=\frac{1}{2}\sum_{n=-\infty }^{+\infty }F(\frac{n\pi}{N})\Sa \left ( \frac{w.N}{2}-\frac{n\pi}{2} \right ).
\label{eq:reconstruction_n_pi_over_2}
\end{equation}  
This is a variation of the cardinal Whittaker-Shannon series \cite{Marks}. 
\begin{equation}
F(w)=\sum_{n=-\infty }^{+\infty }F(\frac{2\pi n}{N}).\Sa \left ( \frac{wN}{2}-{n\pi} \right ).
\label{eq:reconstruction_n_pi}
\end{equation}
Observing the series described in Eqn.~\eqref{eq:series}, it is seen that the signal corresponds to a continuous signal defined by samples such that $F\left ( \frac{n\pi}{N} \right )=2I_{|n|}(\beta)$ and 
\begin{equation}
F(w)=2I_{\left | \frac{Nw}{\pi} \right |}(\beta).
\end{equation}
And the spectrum is given by:
\begin{equation}
W_{CIR;1}(w)=\frac{2NI_{\left | \frac{Nw}{\pi} \right |}(\beta)}{e^ {\beta}}.
\end{equation}
In the case of the causal window, $w_{CIR;1}(t)$, the application of the time-shift theorem provides the spectrum:
\begin{equation}
W_{CIR;2}(w)=\frac{2NI_{ \frac{N}{\pi} \left |w \right |}(\beta)}{e^ {\beta}}.e^{-jw\frac{N}{2}}.
\end{equation}
It is worth remembering that the $\nu$ argument of the $I_{\nu}(z)$ is a real number \cite{Abramowitz}.
\section{Spectrum of the Normal Circular Window: the Discrete Case}
The spectrum of $w_{CIR;3}[n]$ (non causal signal) is evaluated by the discrete-time Fourier transform (Eqn.~\eqref{eq:DiscretetimeFOURIER}), 
\begin{equation}
W_{CIR;3}\left ( e^{j\omega} \right )= \sum_{n=-N/2}^{N/2}\frac{e^{\beta.\cos \left ( \frac{\pi n}{N} \right ) }}{e^{\beta}}.e^{-j\omega n}.
\end{equation}
It is proposed to use the infinite series
\begin{equation}
e^{\beta \cos \theta}= \sum_{k=-\infty }^{+ \infty }I_{|k|}(\beta).\cos~k\theta.
\end{equation}
yielding:
\begin{equation}
W_{CIR;3}\left ( e^{j\omega} \right )= e^{-\beta} \sum_{k=-\infty }^{+ \infty }I_{|k|}(\beta). 
\sum_{n=-N/2}^{N/2}\cos\left ( n\frac{\pi k}{N} \right )e^{j\omega n}.
\end{equation}
Now, using the DTFT of a pulse $\cos\left ( n\frac{k \pi}{N} \right )$ confined within $|n| \leq N/2$, we take the spectrum of the cosine-tip window with parameter $\alpha=0$.\\
\begin{equation*}
w_{\alpha=0;3}[n]:=\cos \frac{2 \pi}{N}n ~~~~~~-N/2 \leq n \leq N/2,
\end{equation*}
whose spectrum is given in Eqn.~\eqref{eq:alpha0}. By adjusting properly, one gets
\begin{equation}
W_{CIR;3}\left ( e^{j\omega} \right )= \sum_{k=-\infty }^{+\infty}\frac{I_{|k|}(\beta)}{2e^{\beta}}
\left [ \asinc_{N+1}\left ( \omega-\frac{k \pi}{N} \right )+\asinc_{N+1}\left ( \omega+\frac{k \pi}{N} \right )   \right ],
\end{equation}
or, finally
\begin{equation}
W_{CIR;3}\left ( e^{j\omega} \right )= \sum_{k=-\infty }^{\infty}\frac{I_{|k|}(\beta)}{e^{\beta}}.
\asinc_{N+1} \left ( \omega-\frac{k \pi}{N} \right ).
\end{equation}
\begin{figure}[h!]
\centering
\includegraphics[width=9cm,height=5.5cm]{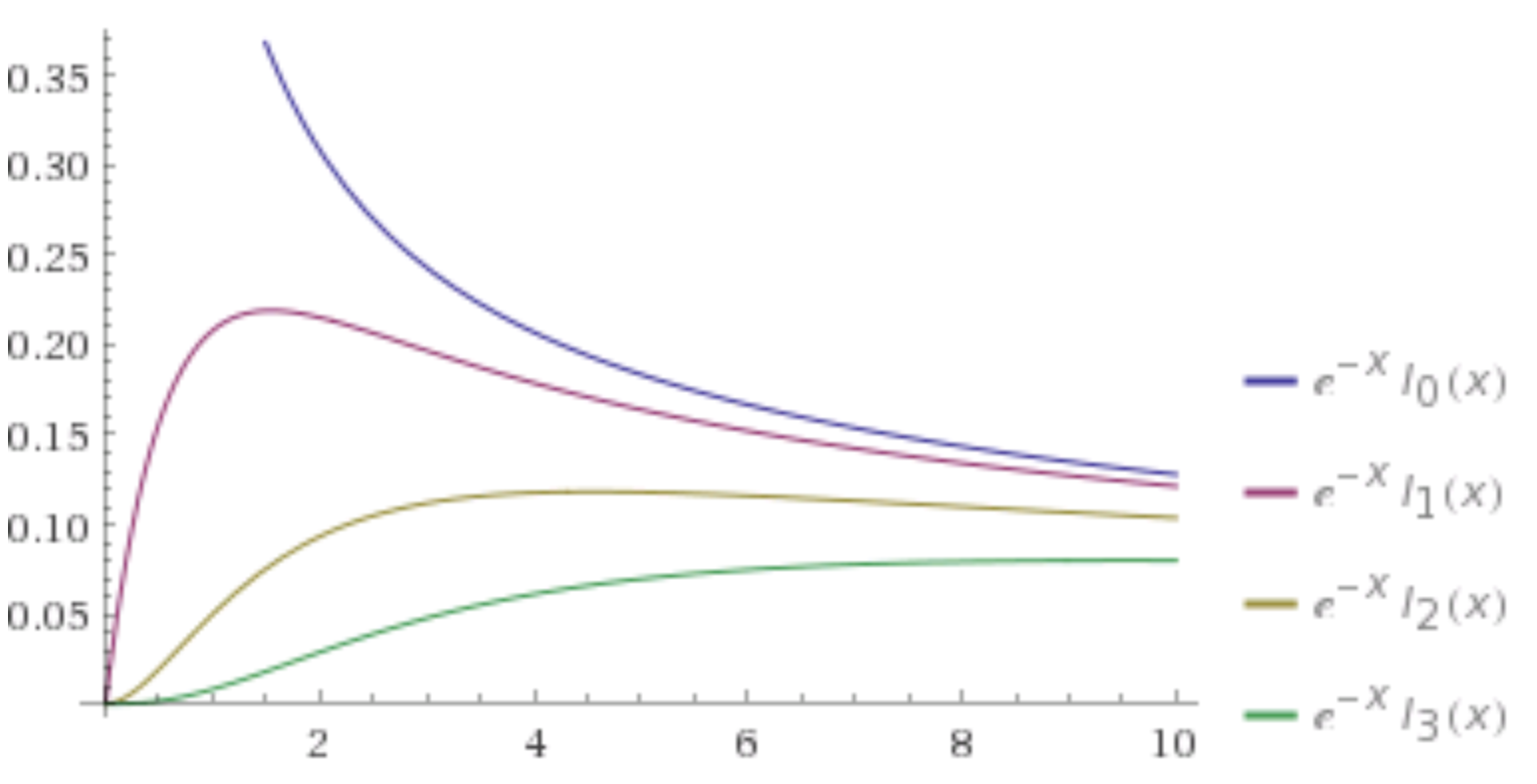}
\caption{\small{Asymptotic behavior of the coefficients $I_{|k|}(\beta)/e^{\beta}$}}.
\label{fig:asymptotic}
\end{figure}
For fixed $\beta$, as $|k|$ increases, the coefficients $I_{|k|}(\beta)/e^{\beta}$ are bounded by $1/\sqrt{2 \pi \beta}$ and decrease monotonically as observed in Figure ~\ref{fig:asymptotic}. Now it is possible go further and evaluate the taper performances according to \cite{Harris}, \cite{Nuttall}, \cite{Geckinli}.

\section{Conclusions}
The closeness to the normal distribution and the fact that it is associated with a shape linked to the maximum entropy for circular data suggests interesting properties to be explored in later investigations. It is expected that the windowing by von Mises' taper can improve spectral evaluation in cases with circular symmetry.

\end{document}